\documentclass[12pt]{article}
\usepackage{graphicx}
\usepackage{mathrsfs}
\usepackage[utf8]{inputenc} 
\usepackage{tikz}
\usepackage{graphicx} 
\usepackage{booktabs} 
\usepackage{array} 
\usepackage{paralist} 
\usepackage{verbatim} 
\usepackage{subfig} 
\usepackage{amssymb} 
\usepackage{amsmath} 
\usepackage{multirow} 
\usepackage{rotating} 
\usepackage[version=3]{mhchem}  
\usepackage[affil-it]{authblk}
\usepackage{mathrsfs}
\usepackage{mathabx}
\usepackage{fancyhdr} 

\newcommand{\bi}[1]{\bf{#1}}

\begin{document}

\title{Decomposition of direct product at an arbitrary Brillouin zone point: \(D^{(\bigstar{R})(m)}\) \(\otimes\) \(D^{(\bigstar{-R})(m)}\)}

\author{Jian Li%
  \thanks{\texttt{jianli@sci.ccny.cuny.edu}}, 
Jiufeng J. Tu, Joseph L. Birman}
\affil{Physics Department, The City College of New York,\\
160 Convent Avenue, New York 10031, USA}

\maketitle

\begin{abstract}
A general rule is presented for the decomposition of the direct product of irreducible representation at arbitrary Brillouin zone point \(\bi{R}\) with its negative: the number of the appearences of the zone center representation equals the dimensionality of the representation. This rule is applicable for all space groups. Although in most situations the interesting physics takes place at high symmetry points in the Brillouin zone, this general rule is useful for situations where double excitations are considered. It is shown that double excitations from arbitrary Brillouin point \({\bi R}\) have the right symmetry to participate in all optical experiments regardless of polarization directions.
\end{abstract}

{{\bf PACS:} 61.50.Ah}

\section{Introduction}

In solid state physics, elementary excitations can be classified according to their symmetries. Selection rules based on symmetry considerations have been sucessfully applied to transitions involving initial states and final states. Photons carry negligible wave vector therefore only zone center excitations participate in an optial process at first order. Second order double excitations, on the other hand, are also symmetry allowed to participate in optical transitions and sometimes even dominate the first order single excitation~\cite{Tanabe}. Due to the possibly high density of states at high symmetry points~\cite{Kudryavtseva}, double excitations are only considered at high symmetry points, lines and surfaces. These special, high symmetry points, lines or surfaces are vanishingly small in number compared to arbitrary point \(\bi{R}\) which occupies almost the entire Brillouin zone. Second order double excitations from \(\bi{R}\) can also participate in experiments therefore it is interesting from a physics point of view that the direct product of \(D^{(\bigstar{R})(m)}\) \(\otimes\) \(D^{(\bigstar{-R})(m)}\) is performed.

When doing the decomposition of direct product \(D^{(\bigstar{R})(m)}\) \(\otimes\) \(D^{(\bigstar{-R})(m)}\), a general rule is found: the reduction coefficients of the \(\Gamma\) point representations equals the dimensionality of the representation (see equation~\ref{eqn:test}). We will show below that this rule holds for all kinds of space groups: Fedorov groups, layer groups, rod groups, wallpaper groups, frieze groups and line groups. The rule presented in this note saves the labor in determining the coefficients in the standard linear algebraic method. To our knownledge no such analysis exist in the literature.

This note is arranged as follows: theoretical background on both the irreducible representations and reduction coefficients of direct product is given in section II, where the linear algebraic method is introduced; the calculations of reduction coefficients of an arbitrary point with its negative are represented in section III and a general rule is proved; discussions are given in section IV and \(\bi{R}\) point of \(O_{h}^{5}\) is used to demonstrate the helpfulness of our finding.

\section{Theoretical background}

The irreducible representations of group \(\mathscr{G}\) is a set of matrices \(D^{(\bigstar{k})(m)}(\{\varphi|t(\varphi)\})\) that has a one-to-one correspondence with group element \(\{\varphi|\bi {t(\varphi)}\}\) such that \(D^{(\bigstar{k})(m)}(\{\varphi_{1}|t(\varphi_{1})\})\) \(\cdot\) \(D^{(\bigstar{k})(m)}(\{\varphi_{2}|t(\varphi_{2})\})\) = \(D^{(\bigstar{k})(m)}(\{\varphi_{1}|t(\varphi_{1})\} \cdot \{\varphi_{2}|t(\varphi_{2})\})\). \(\bi{k}\) is the label for Brillouin zone point and \(m\) is for the \(m\)th irreducible representation at \(\bi{k}\). Rotational operations bring wave functions with wave vector \(\bi k\) to wave functions with wave vectors \(\varphi \cdot \bi k\). Some of these wave vectors are equivalent to \(\bi k\): \(\varphi_{i} \cdot \bi k\) = \(\bi k\) + \(\bi {B_{H}}\) and those symmetry operations define \(\mathscr{G}_{k}\), {\em the group of wave vector} . The inequivalent set of \(\{\varphi \cdot \bi k\}\) is called the {\em star} of \(\bi k\) and each inequivalent one of \(\{\varphi \cdot k\}\) is an {\em arm} of the star.

In most cases, the representation of wave vector group \(\mathscr{G}_{k}\) can be obtained from their corresponding point groups \(D^{({k})(m)}(\{\varphi|t(\varphi)\})\) = \(D^{(k)(m)}(\varphi)\) \(\cdot\) \(e^{-{\rm i}k \cdot t(\varphi)}\) where \(D^{(k)(m)}(\varphi)\) is the representation of the corresponding point group. For example, at \(\Gamma\) point, the representation of space group is simply the representation of the point group: \(D^{(\Gamma)({i})}(\{\varphi|t\})\) = \(D^{(\Gamma)({i})}(\varphi)\). The exceptions are the Brillouin zone points on the boundary of the non-symmorphic space groups. Their representations need special treatment~\cite{Birman}. and the results have already been tabulated by Kovalev~\cite{Kovalev} and Zak~\cite{Zak}. Irreducible representation \(D^{(\bigstar{k})(m)}\) of \(\mathscr{G}\) can be induced from \(D^{({k})(m)}\) of \(\mathscr{G}_{k}\)
\( D^{(\bigstar{k})(m)}(\{\varphi|t\})_{\sigma \tau} = \dot D^{(k)(m)}(\{\varphi_{\sigma}|t_{\sigma}\}^{-1} \cdot \{\varphi|t\} \cdot \{\varphi_{\tau}|t_{\tau}\})
\) and the {\em dotted matrix} \(\dot D^{(k)(m)}(\{\varphi|t\})\) are defined as:
\begin{equation}
 \dot D^{(k)(m)}(\{\varphi|t\}) = \left\{
  \begin{array}{l l l}
   0& \textrm{if } \{\varphi|t\} \textrm{ is not in } \mathscr{G}_{k} \\
   D^{(k)(m)}(\{\varphi|t\})       & \textrm{if } \{\varphi|t\} \textrm{ is in } \mathscr{G}_{k}.
  \end{array} \right.  \label{eqn:dot}
\end{equation}
\(D^{(\bigstar{k})(m)}\) is in a block structure with \(\sigma\) and \(\tau\) as indices for blocks. \(\varphi_{\sigma}\) and \(\varphi_{\tau}\) are coset representatives of \(\mathscr{G}_{k}\). The character \(\chi^{(\bigstar{k})(m)}(\{\varphi|t\})\) of \(D^{(\bigstar{k})(m)}\) are defined accordingly.

The direct product of two representations is in general reducible and can be decomposed as summation of many irreducible representations. The group theoretical job is to determine the reduction coefficients \((\bigstar{k}m\bigstar{k'}m'|\bigstar{k''}m'')\) which is defined as
\(
 D^{(\bigstar{k} \otimes \bigstar{k'})(m \otimes m')} = \sum_{\bigstar{k''}, m''} (\bigstar{k}m\bigstar{k'}m'|\bigstar{k''}m'')D^{(\bigstar{k''})(m'')} 
\), or equivalently,
\begin{equation}
 \chi^{(\bigstar{k})(m)} \chi^{(\bigstar{k'})(m')}= \sum_{\bigstar{k''}, m''} (\bigstar{k}m\bigstar{k'}m'|\bigstar{k''}m'')\chi^{(\bigstar{k''})(m'')}. \label{eqn:chi1}
\end{equation}
Among the different ways of determining the coefficients \((\bigstar{k}m\bigstar{k'}m'|\bigstar{k''}m'')\), method of linear algebraic equations is most straightforward. Choose as many symmetry elements as needed in equation~\ref{eqn:chi1} and the coefficients \((\bigstar{k}m\bigstar{k'}m'|\bigstar{k''}m'')\) can be obtained once the number of linearly independent equations equals the number of coefficients. But before that, not a single coefficient can be determined. Sometimes the number of coefficients is quite big and the calculation is complicated.

\section{Decomposition of direct product \(D^{(\bigstar{R})(m)}\) \(\otimes\) \(D^{(\bigstar{-R})(m)}\)}

\(\bi R\) is the arbitrary point in the Brillouin zone. Arbitrary means that \(\bi R\) = \((k_{x}, k_{y}, k_{z})\) and the group of wave vector contains only the identity operator: \(\mathscr{G}_{R}\) = \(\{E|0\}\). \(\mathscr{G}_{R}\) has only one one-dimensional representation therefore the superscript \(m\) is dropped from now on. The character of \(D^{(\bigstar{R})}\) is:
\begin{equation}
 \chi^{(\bigstar{R})}(\{\varphi|t\}) = \left\{
  \begin{array}{l l}
   0 & \textrm{if } \varphi \neq E \\
   \sum_{\sigma} e^{-{\rm i}R \cdot (\sigma t)} & \textrm{if } \varphi = E.
  \end{array} \right. \label{eqn:chi3}
\end{equation}
where coset representative \(\varphi_{\sigma}\) is now the whole set of \(n\) rotational operations. Representations contained in \(D^{(\bigstar{R})(m)}\) \(\otimes\) \(D^{(\bigstar{-R})(m)}\) can be devided into three catagories: \(\Gamma\), \(\bigstar{X}\) and \(\bigstar{R'}\): \(\Gamma\) = (0, 0, 0); \(\bigstar{X}\) are some high symmetry points, lines or surfaces besides \(\Gamma\), such as (\(k_{x}, k_{y}, k_{z}\)) + (\(-k_{x}, -k_{y}, k_{z}\)) = (\(0, 0, 2k_{z}\)) and \(\bigstar{R'}\) is arbitrary points such as (\(k_{x}, k_{y}, k_{z}\)) + (\(k_{x}, k_{y}, k_{z}\)) = (\(2k_{x}, 2k_{y}, 2k_{z}\)). There could be more than one \(\bigstar{X}\) and \(\bigstar{R}\) therefore they are denoted as \(\bigstar{X}\), \(\bigstar{X'}\), \(\bigstar{R'}\), \(\bigstar{R''}\)and so on.
\begin{eqnarray}
\chi^{(\bigstar{R})} \otimes \chi^{(\bigstar{-R})} &= & \sum_{i}(\bigstar{R} \bigstar{-R}| \Gamma i) \chi^{(\Gamma)({i})} \nonumber \\
&& + \sum_{\bigstar{X}, i}(\bigstar{R} \bigstar{-R}| \bigstar{X} i) \chi^{(\bigstar{X})({i})}  \nonumber \\
&&+ \sum_{\bigstar{R'}}(\bigstar{R} \bigstar{-R}| \bigstar{R'}) \chi^{(\bigstar{R'})} \label{eqn:chi3}
\end{eqnarray}
We want only \((\bigstar{R} \bigstar{-R}| \Gamma i)\). For \(\{\varphi|t\}\) = \(\{E|0\}\), equation~\ref{eqn:chi3} reads:
\begin{eqnarray}
n \times n &= & \sum_{i}(\bigstar{R} \bigstar{-R}| \Gamma i) \chi^{(\Gamma)({i})}(\{E|0\}) \nonumber \\
&& + \sum_{\bigstar{X}, i}(\bigstar{R} \bigstar{-R}| \bigstar{X} i) \chi^{(\bigstar{X})({i})}(\{E|0\})  \nonumber \\
&&+ \sum_{\bigstar{R'}}(\bigstar{R} \bigstar{-R}| \bigstar{R'}) \chi^{(\bigstar{R'})}(\{E|0\}) \label{eqn:E1}
\end{eqnarray}
However, the total dimensionality of \(\Gamma\) point representations in the direct product equal the number of zeros in \(\bigstar{R}\) \(\otimes\) \(\bigstar{-R}\). Each arm \(\varphi\) \(\cdot\) \(R\) and its negative \(\varphi\) \(\cdot\) \(-R\) gives one zone center dimension. The total dimenionality of zone center representations equals the number of arms in \(\bigstar{R}\):
\begin{equation}
n =  \sum_{i}(\bigstar{R} \bigstar{-R}| \Gamma i) \chi^{(\Gamma)({i})}(\{E|0\}) = \sum_{i}(\bigstar{R} \bigstar{-R}| \Gamma i) \chi^{(\Gamma)({i})}(E) \label{eqn:E2}
\end{equation}
For \(\{\varphi|t\}\), \(\varphi\) \(\neq\) \(E\), equation~\ref{eqn:chi3} gives:
\begin{eqnarray}
0&= & \sum_{i}(\bigstar{R} \bigstar{-R}| \Gamma i) \chi^{(\Gamma)({i})}(\{\varphi|t\}) \nonumber \\
&& + \sum_{\bigstar{X}, i}(\bigstar{R} \bigstar{-R}| \bigstar{X} i) \chi^{(\bigstar{x})({i})}(\{\varphi|t\}) \label{eqn:phi1}
\end{eqnarray}
This equation can be simplified using the fact that \(\chi^{(\Gamma)({i})}(\{\varphi|t\})\) are not functions of \(\bi R\) = (\(k_{x}, k_{y}, k_{z}\)) while \(\chi^{(\bigstar{x})({i})}(\{\varphi|t\})\) are. Equation~\ref{eqn:phi1} now becomes:
\begin{equation}
0 =  \sum_{i}(\bigstar{R} \bigstar{-R}| \Gamma i) \chi^{(\Gamma)({i})}(\{\varphi|t\}) = \sum_{i}(\bigstar{R} \bigstar{-R}| \Gamma i) \chi^{(\Gamma)({i})}(\varphi) \label{eqn:phi2}
\end{equation}
Equation~\ref{eqn:phi2} is a set of (\(n-1\)) equations. Combined with equation~\ref{eqn:E2}, the coefficients \((\bigstar{R} \bigstar{-R}|\Gamma i)\) are determined:
\begin{equation}
(\bigstar{R} \bigstar{-R}|\Gamma i) = \chi^{(\Gamma)(i)}(E) \label{eqn:result}
\end{equation}
The last step is exactly the same as the orthogonality theorem of the 32 point groups~\cite{Hamermesh}. It shows that the number of appearances of any zone center representation in the direct product of \(D^{(\bigstar{R})(m)}\) \(\otimes\) \(D^{(\bigstar{-R})(m)}\) equals the dimensionality of that representation in the corresponding point group. The above analysis applies to both symmorphic and non-symmorphic space groups. We want to emphasis that the above analysis is not restricted to Fedorov space groups. The same analysis applies to space group of different dimensions: Fedorov groups, layer groups, rod groups, wallpaper groups, frieze groups and line groups.

\section{Discussion}
\begin{figure}
\begin{center}
\includegraphics[width = 6cm]{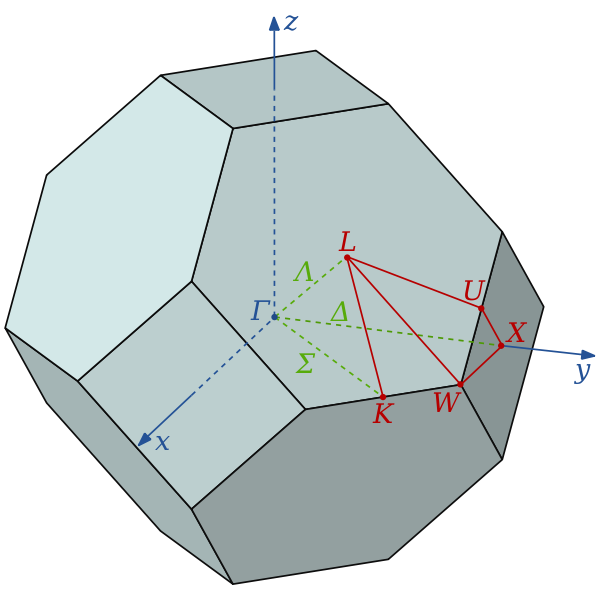} 
\caption{The first Brillouin zone of face-centered-cubic crystal is a truncated octahedron. Symmetry labels for high symmetry lines and points are given.} \label{fig:FCC}
\end{center}
\end{figure}
Before discussing the physical interpretations of the general rule, an example is given. The decomposition of direct product of an arbitrary Brillouin point \(\bigstar{R}\) in \(O_{h}^{5}\) (see Fig.~\ref{fig:FCC}) space group with its negative \(\bigstar{-R}\) is: 
\begin{eqnarray}
D^{(\bigstar{R})(1)}\otimes D^{(\bigstar{-R})(1)}= D^{(\Gamma)(1+)}+D^{(\Gamma)(2+)}+2D^{(\Gamma)(3+)}+3D^{(\Gamma)(4+)}\nonumber \\
+3D^{(\Gamma)(5+)}+D^{(\Gamma)(1-)}+D^{(\Gamma)(2-)}+2D^{(\Gamma)(3-)}+3D^{(\Gamma)(4-)} +3D^{(\Gamma)(5-)} \label{eqn:test}
\end{eqnarray}
plus many other terms that are not at \(\Gamma\) point. The direct product of \(D^{(\bigstar{R})(m)}\) \(\otimes\) \(D^{(\bigstar{-R})(m)}\) in \(O_{h}^{5}\) contains 34 stars. Obviously the full determination of the 34 coefficients using the linear algebraic method is rather tedious and one must go to programs for help. {\em Bilbao Crystallographic Server}~\cite{Bilbao1, Bilbao2, Bilbao3} calculates reduction coefficients of any direct product. The result for \({\bi R}\) of \(O_{h}^{5}\) agrees with our predict. Similar tests has been carried out at different (arbitrary) Brillouin zone points of different space groups (especially \(O_{h}\) and \(D_{6h}\)) and equation~\ref{eqn:result} is proved to be right.

It is seen from equation~\ref{eqn:result} that all the zone center representations are contained in the direct product of \(D^{(\bigstar{k})(m)}\) and \(D^{(\bigstar{-k})(m)}\). As a result, double excitations from \(\bi R\) point has the right symmetry to participate in optical transitions regardless of photon polarization directions. Although, in general, the density of states of excitations (such as phonons and magnons) is low at \(\bi R\) point due to the absence of Van Hove singularities, the contributions from \(\bi R\) point which occupies almost the entire Brillouin zone could contribute to the background of the spectrum in optical experiments.

To summerize, a general rule on the decomposition of \(D^{(\bigstar{R})(m)}\) \(\otimes\) \(D^{(\bigstar{-R})(m)}\) is presented and proved. It applies to Fedorov groups, layer groups, rod groups, wallpaper groups, frieze groups and line groups. The physical meaning is that from symmetry point of view double excitations (phonons, magnons) from the vast majority of the Brillouin zone \(\bi R\) could contribute to the backgroud of optical experiments regardless of photon polarizations.

\end{document}